\documentclass[prd,showpacs,superscriptaddress,twocolumn,floatfix,preprintnumbers,altaffilletter]{revtex4-1}
\usepackage{booktabs}
\usepackage{amssymb, amsmath}
\usepackage[utf8]{inputenc}
\usepackage{epsfig}
\usepackage{xcolor}
\usepackage{graphicx}
\usepackage[colorlinks=true]{hyperref}
\usepackage[caption=false]{subfig}
\usepackage{xspace}

\usepackage{ifthen}
\usepackage{etoolbox}

\usepackage{import}
\usepackage{longtable}
\usepackage{graphicx}
\usepackage{amsmath,amssymb}
\usepackage{color}
\usepackage{units}
\usepackage{epstopdf}
\usepackage{hyperref}
\usepackage{multirow}
\usepackage[caption=false]{subfig}
\usepackage{float}

\usepackage{import}
\usepackage{longtable}
\usepackage{graphicx}
\usepackage{amsmath,amssymb}
\usepackage{color}
\usepackage{units}
\usepackage{epstopdf}
\usepackage{hyperref}
\usepackage{multirow}
\usepackage[caption=false]{subfig}
\usepackage{float}
\newcommand{\fixme}[1]{{\color{red}{#1}}}

\newcommand{\beq}{\begin{equation}}
\newcommand{\eeq}{\end{equation}}
\newcommand{\bdm}{\begin{displaymath}}
\newcommand{\edm}{\end{displaymath}}

\definecolor{Gray}{gray}{0.9}
\definecolor{orange}{rgb}{0.9,0.5,0}

\newcommand*\diff{\mathop{}\!\mathrm{d}}

\newcommand{\icecube}[0]{IceCube}

\newcommand{\Msun}{\>{\rm M_{\odot}}}

\def\newacronym#1#2#3{\gdef#1{#3 (#2)\gdef#1{#2}}}

\def\cqg{Classical Quantum Gravity}   %Classical and Quantum Gravity
\def\apjl{Astrophys.~J.~Lett}

\newacronym{\gt}{Georgia Tech}{Georgia Institute of Technology}
\newacronym{\icecube}{IceCube}{IceCube South Pole Neutrino Observatory}
\newacronym{\cta}{CTA}{Cherenkov Telescope Array}
\newacronym{\sdsc}{SDSC}{San Diego Supercomputer Center}
\newacronym{\cra}{CRA}{Center for Relativistic Astrophysics}
\newacronym{\nr}{NR}{numerical relativity}
\newacronym{\ornl}{ORNL}{Oak Ridge National Laboratory}
\newacronym{\lisa}{LISA}{Laser Interferometer Space Antenna}
\newacronym{\hst}{HST}{Hubble Space Telescope}
\newacronym{\jwst}{JWST}{James Webb Space Telescope}
\newacronym{\alma}{ALMA}{Atacama Large Millimeter Array}
\newacronym{\ligo}{LIGO}{Laser Interferometer Gravitational Wave Observatory}
\newacronym{\sph}{SPH}{smooth particle hydrodynamics}
\newacronym{\tsi}{TSI}{Terascale Supernova Initiative}
\newacronym{\wmap}{WMAP}{the Wilkinson Microwave Anisotropy Probe}
\newacronym{\cmb}{CMB}{cosmic microwave background}
\newacronym{\ibbh}{IBBH}{intermediate binary black hole}
\newacronym{\grb}{GRB}{Gamma-ray burst}
\newacronym{\grbs}{GRBs}{Gamma-ray bursts}
\newacronym{\hpc}{HPC}{High Performance Computing}
\newacronym{\htc}{HTC}{High Throughput Computing}
\newacronym{\bssn}{BSSN}{Baumgarte-Shapiro-Shibata-Nakamura}
\newacronym{\amr}{AMR}{adaptive mesh refinement}
\newacronym{\mpi}{MPI}{Message Passing Interface}
\newacronym{\cuda}{CUDA}{Compute Unified Device Architecture}
\newacronym{\gui}{GUI}{Graphical User Inferface}
\newacronym{\oit}{OIT}{Office of Information Technologies}
\newacronym{\lbg}{LBGs}{Lyman break galaxies}
\newacronym{\mhd}{MHD}{magneto-hydrodynamics}
\newacronym{\imf}{IMF}{initial mass function}
\newacronym{\sms}{SMS}{supermassive star}
\newacronym{\posix}{POSIX}{Portable Operating System Interface}
\newacronym{\ssd}{SSD}{Solid State Drive}
\newacronym{\os}{OS}{Operating System}
\newacronym{\pace}{PACE}{Partnership for an Advanced Computing Environment}
\newacronym{\oit}{OIT}{Office of Information Technology}
\newacronym{\gtswd}{GTSWD}{Georgia Tech Software Distribution}
\newacronym{\uga}{UGA}{University of Georgia}
\newacronym{\mahpic}{MaHPIC}{Malaria Host-Pathogen Interaction Center}
\newacronym{\nih}{NIH}{National Institutes of Health}
\newacronym{\cdc}{CDC}{Centers for Disease Control and Prevention}
\newacronym{\xsede}{XSEDE}{Extreme Science and Engineering Discovery Environment}
\newacronym{\osg}{OSG}{Open Science Grid}
\newacronym{\egi}{EGI}{European Grid Infrastructure}
\newacronym{\vo} {VO}{Virtual Organization}
\newacronym{\sox}{SoX}{Southern Crossroads}
\newacronym{\slr}{SLR}{Southern Light Rail}
\newacronym{\ddn}{DDN}{DataDirect Networks}
\newacronym{\rsv}{RSV}{Resource and Service Validation}
\newacronym{\cvmfs}{CVMFS}{CernVM File System}
\newacronym{\nfs}{NFS}{Network File Sytem}
\newacronym{\lsc}{LSC}{LIGO Scientific Collaboration}

\newacronym{\DA}{DA}{data analysis}
\newacronym{\CBC}{CBC}{Compact Binary Coalescences}
\newacronym{\lvc}{LVC}{LIGO-Virgo Collaboration}
\newacronym{\aligo}{aLIGO}{Advanced LIGO}
\newacronym{\ldr}{LDR}{LIGO Data Replicator}

\def\si#1{Senior Investigator#1 (SI#1)\gdef\si{SI}}
\def\emri#1{Extreme Mass-Ratio Inspiral#1 (EMRI#1)\gdef\emri{EMRI}}
\def\imbh#1{intermediate mass black hole#1 (IMBH#1)\gdef\imbh{IMBH}}
\def\smbh#1{supermassive black hole#1 (SMBH#1)\gdef\smbh{SMBH}}

\def\bbh#1{binary black hole#1 (BBH#1)\gdef\bbh{BBH}}
\def\bh#1{black hole#1 (BH#1)\gdef\bh{BH}}
\def\ns#1{neutron star#1 (NS#1)\gdef\ns{NS}}
\def\bns#1{binary neutron star#1 (BNS#1)\gdef\bns{BNS}}
\def\nsbh#1{neutron star--black hole#1 (NSBH#1)\gdef\nsbh{NSBH}}
\def\hmns#1{hypermassive neutron star#1 (HMNS#1)\gdef\hmns{HMNS}}
\def\gw#1{gravitational wave#1 (GW#1)\gdef\gw{GW}}
\def\grb#1{Gamma-ray burst#1 (GRB#1)\gdef\grb{GRB}}
\def\pnw#1{post-Newtonian#1 (PN#1)\gdef\pnw{PN}}
\def\eos#1{equation of state#1 (EOS#1)\gdef\eos{EOS}}
\def\gpu#1{graphics processing unit#1 (GPU#1)\gdef\gpu{GPU}}
\def\sn#1{supernova#1 (SN#1)\gdef\sn{SN}}
\def\lae#1{Lyman Alpha Emitter#1 (LAE#1)\gdef\lae{LAE}}
\def\sfr#1{star formation rate#1 (SFR#1)\gdef\sfr{SFR}}
\def\sed#1{spectral energy density#1 (SED#1)\gdef\sed{SED}}
\def\lsst#1{Large Synaptic Survey Telescope#1 (LSST#1)\gdef\lsst{LSST}}
\def\rproc#1{rapid neutron capture#1 ($r$-process#1)\gdef\rproc{$r$-process#1}}

\newacronym{\grhd}{GRHD}{general relativistic hydrodynamics}
\newacronym{\grmhd}{GRMHD}{general relativistic magnetohydrodynamics}
\newacronym{\gw}{GW}{gravitational wave}
\newacronym{\EM}{EM}{electromagnetic}
\newacronym{\utb}{UTB}{University of Texas (Brownsville)}
\newacronym{\aps}{APS}{American Physical Society}
\newacronym{\lsu}{LSU}{Louisiana State University}
\newacronym{\rit}{RIT}{Rochester Institute of Technology}

\graphicspath{{./plots/}}

\newcommand{\macro}[1]{\textcolor{black}{#1}}
\newcommand{\othreeduration}{\macro{204.4}}

\newcommand{\sensfact}{\macro{1.8}}

\newtoggle{endauthorlist}
\toggletrue{endauthorlist}

\begin{document}
\title{All-sky search for long-duration gravitational-wave bursts in the third Advanced LIGO and Advanced Virgo run}

\let\mymaketitle\maketitle
\let\myauthor\author
\let\myaffiliation\affiliation
\author{LIGO Scientific Collaboration, Virgo Collaboration, and KAGRA Collaboration}

\date[\relax]{compiled \today}

\begin{abstract}
After the detection of gravitational waves from compact binary coalescences,
the search for transient gravitational-wave signals with less well-defined waveforms for which matched filtering is not well-suited is
one of the frontiers for gravitational-wave astronomy.
Broadly classified into ``short'' $ \lesssim 1~$\,s and ``long'' $
\gtrsim 1~$\,s duration signals, these signals are expected from a variety of astrophysical processes, including non-axisymmetric deformations
in magnetars or eccentric binary black hole coalescences.  In this
work, we present a search for long-duration gravitational-wave
transients from Advanced LIGO and Advanced Virgo's third observing run
from April 2019 to March 2020.  For this search, we use minimal assumptions for the sky location, event time, waveform morphology, and duration of the source. The search covers the range of $2~\text{--}~ 500$~s in duration and a frequency band of
$24 - 2048$\,Hz.
We find no significant triggers
within this parameter space; we report sensitivity limits on the signal strength of gravitational waves characterized by the root-sum-square amplitude $h_{\mathrm{rss}}$ as a function of
waveform morphology. These $h_{\mathrm{rss}}$ limits improve upon the results from the second observing
run by an average factor of \fixme{\sensfact}.
\end{abstract}

\maketitle

\section{Introduction}

The third observing run of the Advanced LIGO~\citep{aLIGO} and
Advanced Virgo~\citep{adVirgo} detectors has revealed a large number of new gravitational-wave signals from the collision of compact objects. Many binary black hole systems~\citep{O3aCatalog} have been identified. These include GW190521 \cite{AbEA2020b} with the largest progenitor masses discovered so far, and GW190814, a merger containing an object in the ``mass-gap'' between neutron stars and black holes \cite{AbEA2020c}. A second binary neutron star (BNS) system was also discovered, GW190425 \citep{AbEA2020}, following the 
first BNS system GW170817~\citep{AbEA2017b}, which
also produced GRB 170817A~\citep{GRBGWPaper} and
an optical transient, AT 2017gfo~\citep{MMAPaper}.
In addition, two neutron star-black hole (NSBH) binary coalescences
(GW200105\_162426 and GW200115\_042309) have been finally detected
\cite{AbEA2021a}.

\begin{figure*}[hbtp!]
  {\mbox{\includegraphics[width=5.0in]{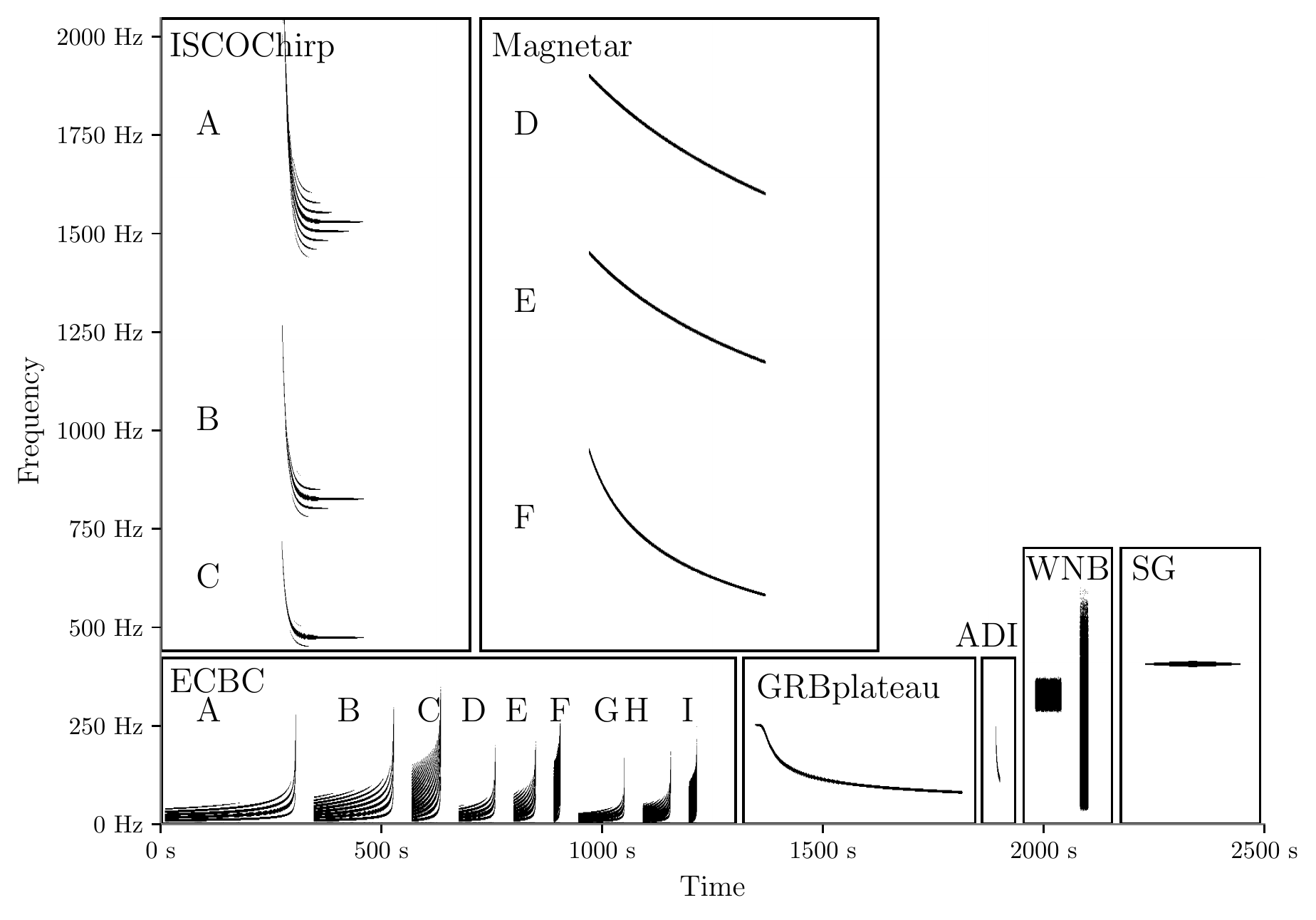}}}
\caption{Time-frequency spectrogram of the reference waveforms used in this search. The ISCOChirp waveforms have been shifted up in frequency by 50 Hz for readability. Durations range from 6\,s (ADI-B) to 470 \,s (GRBplateau). 
}
\label{fig:waveformtraces}
\end{figure*}

Searches for ``long'' $ \gtrsim 1~$\,s duration signals cover a
variety of astrophysical phenomena~\citep{ThKa2011}. 
While well-modeled compact binary coalescences can have similar durations in the sensitive band of the interferometers and the methods employed in this paper are also sensitive to them, this search is not aimed at these systems as matched filtering is much more sensitive.
However, there are less well-defined waveforms for which matched filtering is not well-suited.
Plausible processes include fallback accretion onto a rapidly rotating black
hole~\citep{VP2016} or in newborn neutron
stars~\citep{Lai1995,PiOt2011,PiTh2012}.
They also include non-axisymmetric deformations
in magnetars~\citep{CoMe2009} or accretion disk instabilities and
fragmentation of material spiraling into a black
hole~\citep{PiPf2007,VP2001,VP2008} and in the central engine
of super-luminous supernovae \citep{VP2017,VP2019}.
Figure~\ref{fig:waveformtraces} shows several different realizations of the corresponding waveform morphologies.

In this paper, we present the results of unmodeled long-duration transient searches from the third observing run, updating the results
from the first two observing runs~\citep{AbEA2017g,AbEA2019}.  
As in previous analyses~\citep{AbEA2017g,AbEA2019,AbEA2017k,AbEA2018b}, three
pipelines are used; their different assumptions and data handling techniques yield complementary
coverage of the signal models.

The paper is organized as follows. The data used in the analysis is described in Section~\ref{sec:dataset}.
The algorithms used to analyze the data are outlined in
Section~\ref{sec:searches}.  The results of the analysis and their
implications are discussed in Section~\ref{sec:results}.

\section{Data}
\label{sec:dataset}

The third observing run (O3) of Advanced LIGO and Advanced Virgo spanned April
1, 2019 - March 27, 2020.
O3 was broken up into two segments, with O3a running April 1, 2019 - Oct 1, 2019 and O3b running November 1, 2019 - March 27, 2020; together, these correspond to 330~days.
It is customary to assess detector
sensitivities in terms of a binary neutron star inspiral range (BNS range), which is the
average distance to which these signals could be detected
\cite{AbEA2018c,Chen2021}. Detector upgrades to the LIGO detectors in Hanford,
WA and Livingston, LA yielded binary neutron star ranges of $\sim$115\,Mpc and 133\,Mpc respectively, amounting to improvements of $\sim$50\% with respect to O2.
Similarly, Advanced Virgo reached a binary
neutron star range of $\sim$50\,Mpc, a $\sim$100\% improvement. 
In the following, the algorithms employed require at least two
detectors to be available to process the data; therefore, only
data where both LIGO detectors are simultaneously available is
used. Due to the significant difference in detector alignment and sensitivities, the
Virgo data  in the analysis would not improve the coincidence
selection when the other two detectors are active, while the high rate
of non-Gaussian noise would increase the overall false-alarm rate.
We plan to include Virgo in the analysis of the next observing run.

A major challenge in searches for gravitational-wave transients is non-Gaussian noise. Known sources of noise, including non-linear
sources such as time-varying spectral lines, from, e.g., machinery
on-site, side-bands from the 60 Hz power lines, and time-varying
spectral lines, can be witnessed and subtracted using both linear
Wiener filters~\citep{Davis:2018yrz} and machine learning techniques
\citep{OrCo2020,VaHu2019}. 
The analyses that follow use data for which some of the identified sources of noise that couple in linearly to the detector have been subtracted.
Beyond spectral features, there are transient noise triggers
known as \textit{glitches}, which have a variety of origins
\cite{ZeCo2017}.
Glitch rejection procedures rely on correlations with auxiliary channels~\citep{Aasi:2012wd,Abbott:2016ctn}; yet, noise transients not witnessed by auxiliary sensors remain and reduce sensitivity of the searches~\citep{AbEA2018,DaAr2021}.
Each pipeline, described in the next section, implements different strategies to reduce the impact from glitches.
Altogether, during the third
observing run, coincident data of sufficient quality to be analyzed
totaled \othreeduration~days. Since some time segments are too short to be processed by search pipelines, a small fraction ($<2\%$) of this coincident data is not analyzed.

\section{Searches}
\label{sec:searches}

Long-duration unmodeled searches are now briefly reviewed, and we refer the reader to previous publications for further detail~\citep{AbEA2017g,AbEA2019}.  Most unmodeled searches use time-frequency spectrograms with statistics derived from Fourier transforms or wavelet analysis performed
on consecutive time segments.  
Pattern-recognition algorithms then are
employed to search for gravitational waves in these spectrograms.
These algorithms can be classified as:
``seed-based''~\citep{KhCh2009,Pr2016}, for which pixels above pre-determined
thresholds are clustered, and ``seedless''~\citep{ThCo2013,ThCo2014}, for which
sequences of pixels are derived from generic models, such as B\'ezier
curves~\citep{Far1996,ThCo2013,ThCo2014,CoMe2015b,ThCo2015}. Seedless clustering algorithms are sensitive to narrow-band signals at the price of sensitivity to broadband sources, while seed-based algorithms are generally more sensitive to more generic waveform morphologies.
These algorithms identify candidate gravitational-wave events known as \textit{triggers}.
To estimate the background, all pipelines use ``time-slides,'' \citep{WaBi2010a,WaBi2010b}, where detector data is shifted by non-physical time delays and reanalyzed; this procedure is repeated a sufficient number of times such that at least 50 years of coincident live time is analyzed, allowing for a false alarm rate of 1 per 50 years to be estimated.

Three pipelines are deployed in the analysis: two
different versions of the Stochastic Transient Analysis Multi-detector
Pipeline - all sky (STAMP-AS)
pipeline~\citep{ThKa2011,Pr2016,ThCo2015} and the long-duration
configuration of coherent WaveBurst (cWB)~\citep{KlVe2016}.
The cWB pipeline is seed-based while the 
two STAMP-AS algorithms, Zebragard and Lonetrack, use seed-based and seedless
clustering algorithms respectively.
Altogether, the analyses are
sensitive to transients lasting $2~\text{--}~ 500$~s and covering a
frequency band of $24~\text{--}~ 2048$~Hz.
Due to the short duration of binary black hole signals and the weakness of the coalescences containing neutron stars observed during O3 \cite{AbEA2020}, we are not sensitive to and therefore do
not excise any time around known compact binary coalescences.
All false alarm rates reported are per pipeline, with no combination of searches made outside of reporting the most sensitive limit across the parameter space below.

\emph{STAMP-AS}. Spectrograms, with duration 500\,s and frequency band
$24~\text{--}~ 2048$~Hz and a pixel size of \unit[1]{s} $\times$
\unit[1]{Hz}, are derived with cross-power SNR as the statistic computed in the maps. Non-stationary, high-amplitude spectral features are masked to limit their effect on the search.
Zebragard uses cuts on the fraction of SNR per time bin (summing all pixels of the same time index) and the ratio in SNR between
detectors to remove data transients \cite{AbEA2017g}; Lonetrack does not require this cut due to the narrowband assumption.
During a short period of time, a time segment veto that flags periods of instabilities in the high-power laser at
Hanford is applied on Zebragard triggers \cite{DaAr2021}.

\emph{cWB}. The algorithm used by cWB~\citep{KlVe2016} is based on a
maximum likelihood approach applied to the multiresolution
time-frequency representation of the time series of the
detectors' data. Candidate triggers are identified as a cluster if there
is a coherent excess power in the time-frequency pixel representation over the
network data. The search is performed in the frequency range 24 -- 2048
Hz.
Selection criteria are applied on the duration and on the coherence of the trigger; the coherence coefficient, measuring the degree of correlation between the detectors, must be larger than $0.6$~\citep{KlVe2016}. 
Moreover, the trigger energy-weighted duration, defined as

\begin{equation*}
 d=\sqrt{\frac{\sum{w_i(t-t^{*} )}^2}{\sum{w_i}}} \quad ,
\end{equation*} 
where $t$ is the central time of the pixel, $w$ the energy of the pixel, $t^{*}$ the mean time and the sum is computed over the selected pixels of the event in all the resolutions,
is required to be greater than 1.5 s.
Since observed glitch excess in the $16-48$
Hz band, associated with elevated anthropogenic noise, is different
between the first and second part of the run, the acceptance criteria
in the latter one have been slightly modified. The triggers
have  an energy-weighted duration larger than $0.5$ s and a total
duration greater than $5$\,s, this to ensure increased acceptance for the
eccentric compact binary waveforms family discussed in the next
section.

\section{Results and Future Prospects}
\label{sec:results}

The detection threshold  is defined to be a false alarm rate lower than
1/50~years (equivalent to $6.3 \times 10^{-10}$\,Hz).  None of the pipelines found triggers consistent with such a
false alarm rate; the most significant triggers, non-overlapping between the different pipelines and consistent with the
background, are listed in Table~\ref{table:loudest}.  The most
significant event reported by the cWB algorithm (statistical
significance $\sim$1.7 $\sigma$, p-value 0.088) shows a time-frequency map composed of two
separated excess power cluster pixels, respectively, at 838 Hz and 861
Hz mean frequency.
This trigger appears to be associated with a random (time) coincidence of pixels belonging to two different non-stationary spectral lines of unknown origin, at 838 Hz (present in H1 and L1) and 861 Hz (present in H1).
The STAMP-AS Zebragard and Lonetrack pipeline triggers are consistent with typical events
identified in the background.

To place these results in context, upper limits are derived on the
gravitational-wave strain amplitude using a set of simulated waveforms
added coherently into detector data.  
Waveforms that span the parameter space in both frequency and
time, as well as a sampling of potential astrophysical models, are used. For the
astrophysical models, 
post-merger magnetars (Magnetar)~\citep{DaGi2015},
black hole accretion disk instabilities (ADI)~\citep{VP2001},
newly formed magnetar powering a gamma-ray burst plateau (GRBplateau)~\citep{CoMe2009},
eccentric inspiral-merger-ringdown compact binary coalescence waveforms (ECBC)~\citep{HuMo2017},
and broadband chirps from innermost stable circular orbit waves around rotating black holes (ISCOchirp)~\citep{VP2016} are used (see Ref.~\citep{vPDV2019} for further developments).
To fill out the parameter space, ``ad-hoc'' waveforms, band-limited white noise burst (WNB) and
sine-Gaussian bursts (SG) are also used.  Their time-frequency spectrograms
are shown in Figure~\ref{fig:waveformtraces}.

\begin{table}
\begin{tabular}{cccccc}
\hline
\hline
Pipeline   & FAR   & p-value & Frequency  & Duration  & Time\\
           & [Hz]  &         & [Hz]       & [s]            & [GPS]    \\
\hline
\hline
cWB        & $1.0 \times 10^{-8}$ & 0.088 & 838-861  & 16  & 1252808855\\
Zebragard  & $5.6 \times 10^{-8}$ & 0.40 & 1650-1769 & 21  & 1244819393\\
Lonetrack  & $1.7 \times 10^{-8}$ & 0.14 & 1510-1937 & 417 & 1253105020\\
\hline
\hline
\end{tabular}
\caption{Properties of the most significant coincident triggers found
  by each of the long-duration transient search pipelines during the
  third observing run. FAR stands for false alarm rate, while the
  p-value is the probability of observing at least 1 noise trigger at
  higher significance than the most significant coincident trigger. 
  }
\label{table:loudest}
\end{table}

The upper limits on the
gravitational-wave strain amplitude are typically reported for
unmodeled searches using the root-sum-square gravitational-wave amplitude at the
Earth, $h_{\mathrm{rss}}$,
\begin{equation}\label{eq:hrss}
h_{\mathrm{rss}} = \sqrt{\int_{-\infty}^{\infty} \left(h^2_+(t) +
  h^2_{\times}(t) \right) \diff t},
\end{equation}
where $h_{+}$ and $h_{\times}$ are the two non-whitened signal polarizations.
Simulations are varied with $h_{\mathrm{rss}}$
and injected
uniformly in time, sky location, polarization angle and the cosine of
the inclination angle of the assumed source.

\begin{figure}[hbtp!]
  \mbox{\includegraphics[width=3.5in]{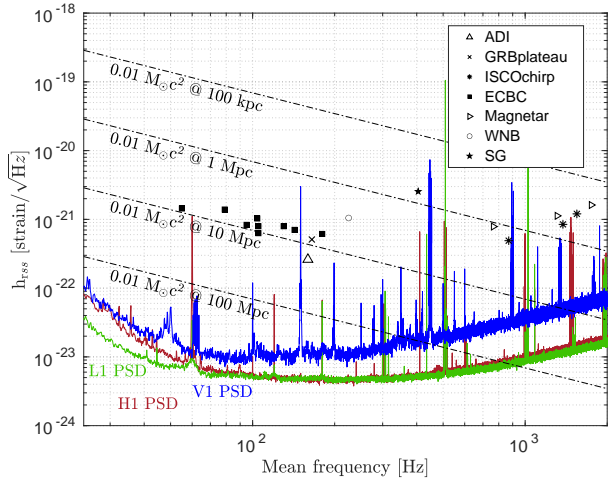}}
\caption{The GW root-sum-square strain amplitude versus mean frequency at 50\%
  detection efficiency and a FAR of 1/50 years.  The red, green and blue
  curves are the averaged amplitude spectral noise densities for Hanford,
  Livingston and Virgo detectors to show that the search results follow the detectors' sensitivity frequency. We also show in dashed-dotted lines the energy of fractions of \protect{$0.01\,\Msun c^2$}, with examples at 100\,kpc, 1\,Mpc, 10\,Mpc and 100\,Mpc shown.}
\label{fig:limits}
\end{figure}

Upper limits on
gravitational-wave strain versus mean frequency for sources detected with
50\% efficiency and a false alarm rate of 1 event in 50 years are
shown in Figure~\ref{fig:limits}. The strongest bounds obtained from the three pipelines are shown on the plot. 
Because each pipeline uses a different clustering algorithm, their relative sensitivities vary
with waveform morphology. Lonetrack, which uses seedless clustering, performs best on magnetar signals (Magnetar and GRBplateau) but is not sensitive to white noise bursts. Zebragard and Coherent WaveBurst give the most constraining values with similar sensitivities for most of the remaining waveforms.
On average, for all waveforms considered in this paper, the $h_{\rm{rss}}$ sensitivity
improved by a factor of \fixme{\sensfact} upon the analysis from the
second observing run~\citep{AbEA2019}.

\begin{figure}[hbtp!]
\includegraphics[width=3.4in]{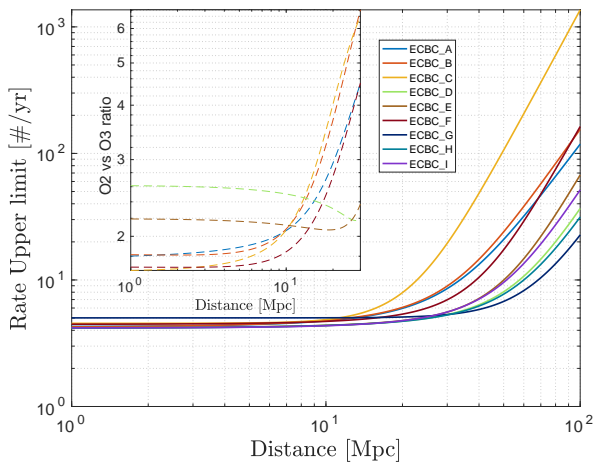}
\caption{Upper limits at $90\%$ confidence level on the rate of
  eccentric compact binary coalescences as a function of the
  distance. Only the best result is shown for each waveform. The inset
  shows the ratio of the rates with respect to O2 results
  \cite{AbEA2019} for ECBC\_A to ECBC\_F (see Table~\ref{table:r90} for parameters).
}
\label{fig:rates}
\end{figure}

For the eccentric binary waveforms, we determine 90\% confidence level limits on the rate of events. We do this using the loudest event statistic method~\citep{BrCr2004}.
Taking as an example the eccentric binary waveforms, the 90\%
upper limits on the event rates as a function of distance are highlighted in
Figure~\ref{fig:rates}.  In addition, Table~\ref{table:r90} gives the upper limits $\mathcal{R}_{90 \%}$ at $90\%$
confidence on the rate of eccentric binary coalescences per unit volume. Following
\cite{2012PhRvD..85l2007A}, and assuming an isotropic and uniform
distribution of sources, $\mathcal{R}_{90 \%}$ is given by
\begin{equation}
\label{eqn:r90}
\mathcal{R}_{90 \%} = \frac{2.3}{4 \pi T \int_0^{r_{\mathrm{max}}}{dr \,r^2 \, \epsilon(r)}},
\end{equation}
where $\epsilon(r)$ is the detection efficiency as a function of
distance, computed as the fraction of transients detectable at a given distance \cite{2012PhRvD..85l2007A}, $r_{\mathrm{max}}$ is the maximum detectable distance, and $T = 204.4$ days is the total observing time. 
For $1.4-1.4$ solar masses eccentric binaries, rate upper limits are $\sim1.5-2$ lower than the ones computed in \cite{nitz2021search} for O2 data. Such improvement can be explained by both the increased sensitivity of the search and the increased livetime between O2 and O3.
For comparison, estimated merger rates from the second LIGO-Virgo GW transient catalogue \cite{theligoscientificcollaboration2021population} are \unit[$23.9^{+14.3}_{-8.6}$]{$\rm{Gpc}^{-3} \rm{yr}^{-1}$} and \unit[$340^{+490}_{-240}$]{$\rm{Gpc}^{-3} \rm{yr}^{-1}$} for binary black holes and binary neutron stars respectively. With eccentric systems expected to be only a fraction of the total binary systems, the upper limits derived are compatible with an absence of detection in this search.

\begin{table}
\begin{tabular}{ccccc}
\hline
 Waveform   & $M_1 [M_{\odot}]$ & $M_2 [M_{\odot}]$& $e$ &   $\mathcal{R}_{90 \%}$ [$\rm{Gpc}^{-3} \rm{yr}^{-1}$] \\
\hline
 ECBC\_A  & $1.4$ & $1.4$ &  $0.2$ &$9.97\times 10^{2}$ \\
 ECBC\_B  & $1.4$ & $1.4$ &  $0.4$ &$8.09\times 10^{2}$ \\
 ECBC\_C  & $1.4$ & $1.4$ &  $0.6$ &$3.21\times 10^{3}$ \\
 ECBC\_D  & $3.0$ & $3.0$ &  $0.2$ &$3.99\times 10^{2}$ \\
 ECBC\_E  & $3.0$ & $3.0$ &  $0.4$ &$8.89\times 10^{2}$ \\
 ECBC\_F  & $3.0$ & $3.0$ &  $0.6$ &$2.43\times 10^{3}$ \\
 ECBC\_G  & $5.0$ & $5.0$ &  $0.2$  &$1.50\times 10^{3}$ \\
 ECBC\_H  & $5.0$ & $5.0$ &  $0.4$ &$5.10\times 10^{2}$ \\
 ECBC\_I  & $5.0$ & $5.0$ &  $0.6$ &$6.98\times 10^{2}$ \\
\hline
\end{tabular}
\caption{Rate upper limits per unit volume at 90\% confidence level on
  eccentric compact binary coalescences with various masses and eccentricity $e$, computed with equation
  \ref{eqn:r90}.}
\label{table:r90}
\end{table}

It is expected that continued improvements both to the gravitational-wave
detectors and to the search algorithms, e.g. \cite{CoCo2016,vPDV2019,PhysRevD.99.123003}, will lead to either detections or
improved limits on this portion of parameter space. Going forward,
increasing the parameter space searched, such as for longer signals, is a high
priority; these signals may include long-lived remnants of binary neutron star mergers, whose detection in gravitational waves may constrain the nature of the remnant \citep{AbEA2018b,VP2016}.
In addition, integration of Advanced Virgo into the analyses
will be important, especially in case of a genuine signal for characterization.
With range improvements of $\sim$50\% expected for
the fourth observing run and more than a factor of 2 expected by the fifth observing run \citep{AbEA2018c},
significant gains in detection possibilities can be expected.

\acknowledgements
This material is based upon work supported by NSF’s LIGO Laboratory which is a major facility
fully funded by the National Science Foundation.
The authors also gratefully acknowledge the support of
the Science and Technology Facilities Council (STFC) of the
United Kingdom, the Max-Planck-Society (MPS), and the State of
Niedersachsen/Germany for support of the construction of Advanced LIGO 
and construction and operation of the GEO600 detector. 
Additional support for Advanced LIGO was provided by the Australian Research Council.
The authors gratefully acknowledge the Italian Istituto Nazionale di Fisica Nucleare (INFN),  
the French Centre National de la Recherche Scientifique (CNRS) and
the Netherlands Organization for Scientific Research, 
for the construction and operation of the Virgo detector
and the creation and support  of the EGO consortium. 
The authors also gratefully acknowledge research support from these agencies as well as by 
the Council of Scientific and Industrial Research of India, 
the Department of Science and Technology, India,
the Science \& Engineering Research Board (SERB), India,
the Ministry of Human Resource Development, India,
the Spanish Agencia Estatal de Investigaci\'on,
the Vicepresid\`encia i Conselleria d'Innovaci\'o, Recerca i Turisme and the Conselleria d'Educaci\'o i Universitat del Govern de les Illes Balears,
the Conselleria d'Innovaci\'o, Universitats, Ci\`encia i Societat Digital de la Generalitat Valenciana and
the CERCA Programme Generalitat de Catalunya, Spain,
the National Science Centre of Poland and the Foundation for Polish Science (FNP),
the Swiss National Science Foundation (SNSF),
the Russian Foundation for Basic Research, 
the Russian Science Foundation,
the European Commission,
the European Regional Development Funds (ERDF),
the Royal Society, 
the Scottish Funding Council, 
the Scottish Universities Physics Alliance, 
the Hungarian Scientific Research Fund (OTKA),
the French Lyon Institute of Origins (LIO),
the Belgian Fonds de la Recherche Scientifique (FRS-FNRS), 
Actions de Recherche Concertées (ARC) and
Fonds Wetenschappelijk Onderzoek – Vlaanderen (FWO), Belgium,
the Paris \^{I}le-de-France Region, 
the National Research, Development and Innovation Office Hungary (NKFIH), 
the National Research Foundation of Korea,
the Natural Science and Engineering Research Council Canada,
Canadian Foundation for Innovation (CFI),
the Brazilian Ministry of Science, Technology, and Innovations,
the International Center for Theoretical Physics South American Institute for Fundamental Research (ICTP-SAIFR), 
the Research Grants Council of Hong Kong,
the National Natural Science Foundation of China (NSFC),
the Leverhulme Trust, 
the Research Corporation, 
the Ministry of Science and Technology (MOST), Taiwan,
the United States Department of Energy,
and
the Kavli Foundation.
The authors gratefully acknowledge the support of the NSF, STFC, INFN and CNRS for provision of computational resources.

\let\author\myauthor
\let\affiliation\myaffiliation
\let\maketitle\mymaketitle
\title{Authors}
\pacs{}

\author{R.~Abbott}
\affiliation{LIGO Laboratory, California Institute of Technology, Pasadena, CA 91125, USA}
\author{T.~D.~Abbott}
\affiliation{Louisiana State University, Baton Rouge, LA 70803, USA}
\author{F.~Acernese}
\affiliation{Dipartimento di Farmacia, Universit\`a di Salerno, I-84084 Fisciano, Salerno, Italy}
\affiliation{INFN, Sezione di Napoli, Complesso Universitario di Monte S. Angelo, I-80126 Napoli, Italy}
\author{K.~Ackley}
\affiliation{OzGrav, School of Physics \& Astronomy, Monash University, Clayton 3800, Victoria, Australia}
\author{C.~Adams}
\affiliation{LIGO Livingston Observatory, Livingston, LA 70754, USA}
\author{N.~Adhikari}
\affiliation{University of Wisconsin-Milwaukee, Milwaukee, WI 53201, USA}
\author{R.~X.~Adhikari}
\affiliation{LIGO Laboratory, California Institute of Technology, Pasadena, CA 91125, USA}
\author{V.~B.~Adya}
\affiliation{OzGrav, Australian National University, Canberra, Australian Capital Territory 0200, Australia}
\author{C.~Affeldt}
\affiliation{Max Planck Institute for Gravitational Physics (Albert Einstein Institute), D-30167 Hannover, Germany}
\affiliation{Leibniz Universit\"at Hannover, D-30167 Hannover, Germany}
\author{D.~Agarwal}
\affiliation{Inter-University Centre for Astronomy and Astrophysics, Pune 411007, India}
% ...

\newpage
\maketitle
\end{document}